\documentclass[a4paper,12pt]{article}
\usepackage{graphicx}
\usepackage[intlimits]{amsmath}
\usepackage{anysize}
\usepackage{subfig}
\usepackage{amssymb}
\marginsize{1 in}{1 in}{0.6 in}{1 in}
\title{Fourier Transform of Electric Signal \\using Kundt's Tube}
\author{Srijit Paul\footnote{The Cyprus Institute and Bergische Universit\"{a}t Wuppertal - 42119, Germany}$~$  and Mahesh Gandikota\footnote{Department of Physics, Syracuse University, NY - 13244, USA}}

\begin{document} 
\maketitle


\begin{abstract}
An experiment to demonstrate the Fourier transform of an electric signal using the Kundt's tube is described. The results of finding the component frequencies and an approximation to the amplitudes of two sinusoidal signals which compose an input electric signal is presented. Undergraduate students are expected to more easily relate to the meaning of a Fourier transform through such mechanical demonstrations.
\end{abstract}

\section{Introduction}

The mathematical technique of Fourier transform has ubiquitous usage in physics. To demonstrate Fourier transform, there are experiments available which use electronic circuits, computer programs\cite{lambert},\cite{web} and bass guitar strings\cite{courtney}. In this paper, we  present a \textit{mechanical} means of physically realizing a Fourier transform. 

A mechanical demonstration of a Fourier transform can be achieved using a Kundt's tube. Kundt's tube is a simple and an easily available apparatus in undergraduate teaching labs where it is  used to measure the speed of sound. In this experiment, we use the resonance property of the Kundt's tube to find the frequencies and amplitudes of component sinusoidal waves which make up an electric signal. 

\section{Theory}
The representation of a function 
as a sum of sine and cosine terms is called a Fourier series\cite{riley}. That is to say, that the right hand side of 
\begin{equation}
 f(x)=\frac{a_0}{2}+\sum_{r=0}^{r=\infty}[a_r\;cos(\omega_r x)+b_r\;sin(\omega_r x)]
\end{equation}
is the Fourier series of the function $f(x)$. $a_0,a_r,b_r$ are called the Fourier coefficients of `component' frequencies $\omega_r$ and can be  mathematically calculated using the orthogonality properties of the above trigonometric functions. 


Historically, Kundt (1866) had used the concept of stationary waves to measure the speed of sound with better accuracy\cite{kundt}.  Even today, we find this tube in undergraduate teaching labs to introduce undergraduate students to one of the methods of measurement of speed of sound.
The Kundt's tube resonates when sound waves of certain frequencies pass through it. For a tube closed at one end and the source at the other end, the resonating frequencies are,
\begin{equation}
 \nu_n=\frac{2n+1}{4}\;\frac{v_s}{L}
\end{equation}
where $L$ is length of tube, $v_s$ is speed of sound in the medium encapsulated by tube and $n\geq 0$ is an integer.



\section{Apparatus}
The setup consists of the Kundt's tube, a cork which is movable along the length of it and a speaker at other end.  Cork dust or thermocol pieces are used as mediums for observing striations in the tube. Open source softwares, \textit{Audacity} and \textit{Praat} are used, for producing input signals and recording the signal from a microphone in the Kundt's tube respectively.
\begin{figure}[h!]
 \centering
 \includegraphics[width=13 cm, height=4 cm]{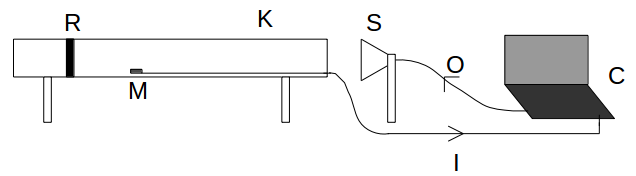}
 \caption{Schematic diagram of setup for the quantitative FT}
\end{figure}
\\\textbf{Legend - }R: Cork, K: Kundt's tube, M: Microphone, S: Speaker, C: Computer, I: Input from microphone, O: Output to speaker

This Kundt's tube is closed only at one end as the air near the speaker-end forms an anti-node forced by the vibrations of speaker's diaphragm. 
\section{Experiment}
\textit{Audacity} is used to add two sinusoidal waves to make an electric signal which is then fed to the speaker. This electric signal is treated as a signal known to be composed of two sinusoidal waves but whose frequencies and amplitudes are unknown. The aim of the experiment is a primitive one: Decompose the electric signal to it's component frequencies and amplitudes. The procedure for achieving the former and latter parts of this aim is presented in sections (\ref{a}) and (\ref{b}) respectively.

\subsection{Qualitative Fourier Transform}\label{a}
The procedure to find the two unknown frequencies of the input signal may be named the qualitative Fourier transform.

\begin{figure}[h!]
 \centering
 \includegraphics[width=12 cm, height=8 cm]{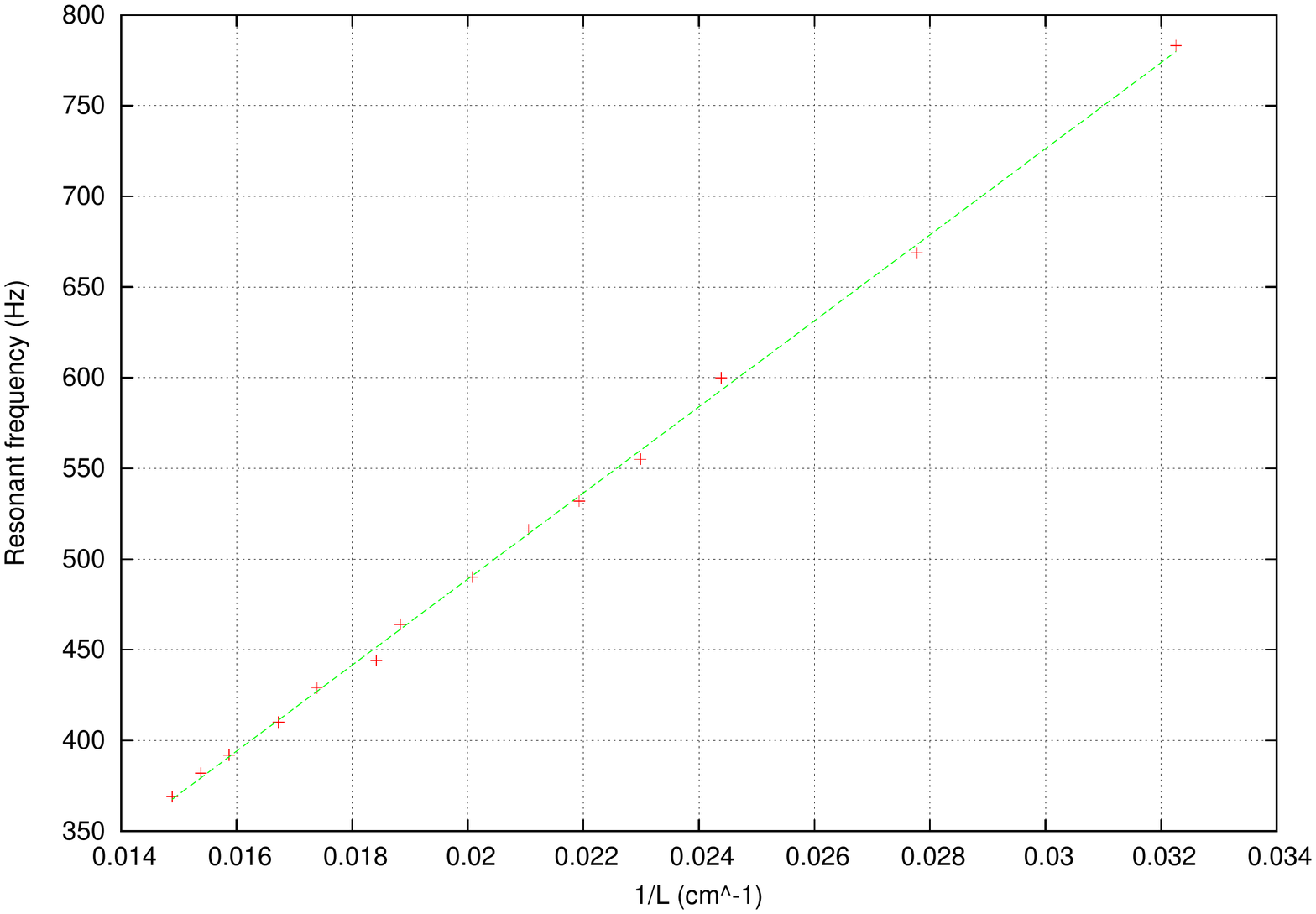}
 \caption{Length-Frequency calibration}\label{c}
\end{figure}

Before finding the component frequencies of the electric signal, a length-frequency calibration is necessary. The Kundt's tube is filled with an optimum amount of cork dust and the length of the tube is fixed by the cork. The resonant frequency for this length is estimated by feeding the speaker a range of frequencies using a function generator and judging which frequency sets up the tallest striations in the cork dust. The resonant frequencies for a certain number of lengths is similarly found and plotted (Fig.(\ref{c})). Note that for the whole calibration, one should decide upon a single mode $n$. We chose $n=1$ and selected frequencies which set up the corresponding pattern in the cork dust. 

Now, the speaker is fed with the unknown signal and the length of the cork is continually changed till the tube resonates at mode $n=1$. This happens at two different lengths\footnote{This is ensured only if the input signal's frequencies are in the range of frequencies covered in the calibration plot (\ref{c}).} which are noted. 
The resonant frequencies corresponding to these lengths is found using the calibration plot (\ref{c}).

This concludes the qualitative Fourier transform as the two unknown frequencies that comprises the signal is found.

In the experiment that we performed, the electric signal fed to the speaker comprised of the frequencies: 358 Hz and 448 Hz and the qualitative FT found resonance at first mode at the lengths of 69 cm and 54 cm corresponding to the frequencies 358 Hz and 448 Hz. 


\subsection{Quantitative Fourier Transform}\label{b}
Quantitative Fourier transform is the procedure to find the amplitudes corresponding to the two unknown frequencies found using the qualitative FT.

To quantify the Fourier transform, we need two more calibration curves. The cork is placed at one of the resonant lengths of the unknown signal and the cork-dust which has served its purpose is removed. A microphone is placed at the displacement anti-node. The speaker is fed with the corresponding resonant frequency at different input voltages using \textit{Audacity} and the average sound intensity measured using \textit{Praat} is recorded for each input voltage.  An Average recorded intensity-input voltage calibration plot is thus obtained.

At this point, when we attempted to verify the shape of the waveform recorded by \textit{Praat}, we found that the waveform was being truncated at the crests and troughs for high input voltages. To remove the associated errors in the average intensity recorded, we scaled down the volume in the computer appropriately. 

\begin{figure}[h!]
 \centering
 \includegraphics[scale=0.4]{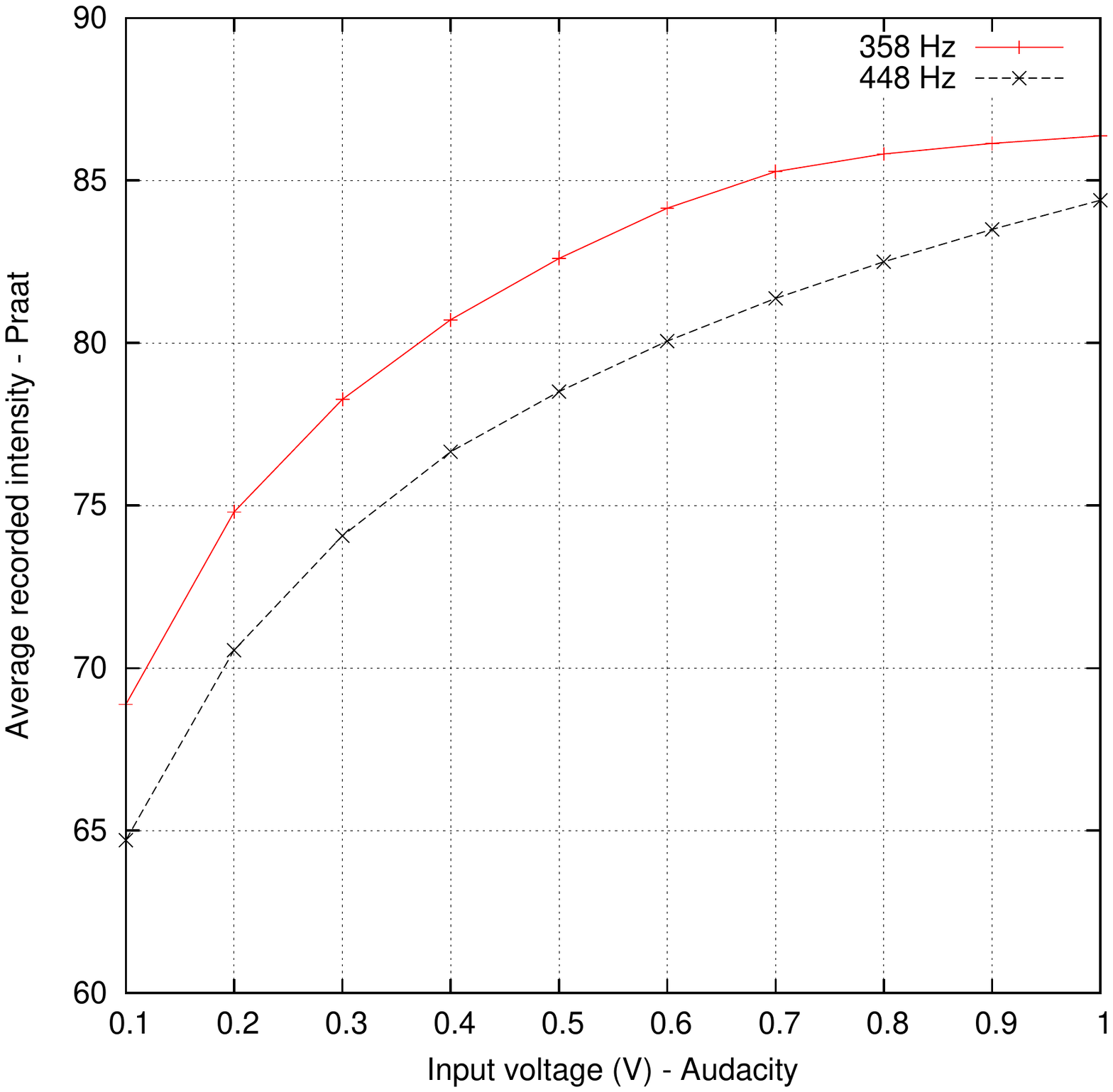}
 \caption{Recorded intensity-input voltage calibration}\label{d}
\end{figure}

Once the calibration for a particular resonant length is done, the same is repeated for the other resonant length. The calibration curves got are shown in Fig.(\ref{d}).

To find the amplitudes of the component frequencies of the electric signal, the Kundt's tube is kept at one of the resonant length and the microphone at the displacement anti-node. The unknown signal is fed to the speaker (at the same scaled down computer volume) and the average intensity is recorded. The corresponding input voltage is found using the calibration plots (\ref{d}). Similar procedure is repeated for the other resonant length. Note that due to the scaling down of computer volume, the amplitudes here correspond to the scaled down electric signal and not the original electric signal which was used in Sec.(\ref{a}).

This concludes the quantitative Fourier transform as the amplitudes of the two component frequencies are estimated. 

In the experiment we performed, we fed these two frequencies at different pairs of amplitudes and the quantitative FT's estimated amplitudes are tabulated in table (\ref{e}).

\begin {table}[htb]
\caption{Quantitative Fourier transform}\label{e}
\begin{center}
\begin{tabular}{|l|l|l|l|l|l|l|l|l|}
\hline
Sl.&Input amp. &Recd. av. &Calib. &Error&Input amp. &Recd. av.&Calib.&error\\
No.&(V) 358 Hz& intensity &reading&&(V) 448 Hz&intensity &reading&\\
\hline
1&0.2 & 75.08 & 0.2 & 0 & 0.4 & 77.08 & 0.42 & 0.02 \\
2&0.4 & 80.58 & 0.39 & 0.01 & 0.2 & 75.53 & 0.35 & 0.15 \\
3&0.5 & 82.49 & 0.49 & 0.01 & 0.5 & 80.26 & 0.61 & 0.11 \\
4&0.5 & 82.39 & 0.49 & 0.01 & 0.6 & 81.24 & 0.69 & 0.09 \\
5&0.5 & 82.25 & 0.48 & 0.02 & 0.7 & 82.16 & 0.77 & 0.07 \\
6&0.5 & 82.09 & 0.47 & 0.03 & 0.8 & 82.85 & 0.83 & 0.03 \\
7&0.6 & 83.88 & 0.58 & 0.02 & 0.5 & 80.8 & 0.65 & 0.15 \\
8&0.7 & 85 & 0.67 & 0.03 & 0.5 & 81.46 & 0.7 & 0.2 \\
9&0.8 & 84.85 & 0.66 & 0.14 & 0.5 & 82.02 & 0.75 & 0.25 \\
10&0.8 & 82.42 & 0.49 & 0.31 & 0.4 & 79.24 & 0.54 & 0.14 \\
11&0.4 & 76.89 & 0.26 & 0.14 & 0.4 & 77.35 & 0.43 & 0.03 \\
12&0.4 & 76.76 & 0.25 & 0.15 & 0.8 & 81.47 & 0.71 & 0.09 \\
\hline
\end{tabular}
\end{center}
\end{table}

It can be observed from table (\ref{e}) that other than readings 8, 9 and 10, the amplitude measurements fall under a resolution of 0.15 V hence giving a rough estimate of the input amplitudes in the unknown electric signal. 

\section{Discussion and Conclusions}
The complete process of finding out the component frequencies of input signal by scanning the length of Kundt's tube and finding out the amplitudes of the corresponding frequencies using calibration curves may be called as performing the Fourier transform of an electric signal using Kundt's tube.

Mechanical systems having discrete resonant frequencies can be considered as candidates for performing a Fourier transform of an electric signal. Preceding the mechanical system should be a device which can convert electric signals to mechanical signals which can be fed to a resonace capable mechanical system. Example - springs of different resonant frequencies put on a hanger. A mechanical vibrator can be used to convert electric signals to mechanically vibrate the hanger. Only those springs whose resonant frequencies match with the frequency of the sinusoidal waves in electric signal will show resonance.

\begin{figure}[h!]
 \centering
 \includegraphics[scale=0.3]{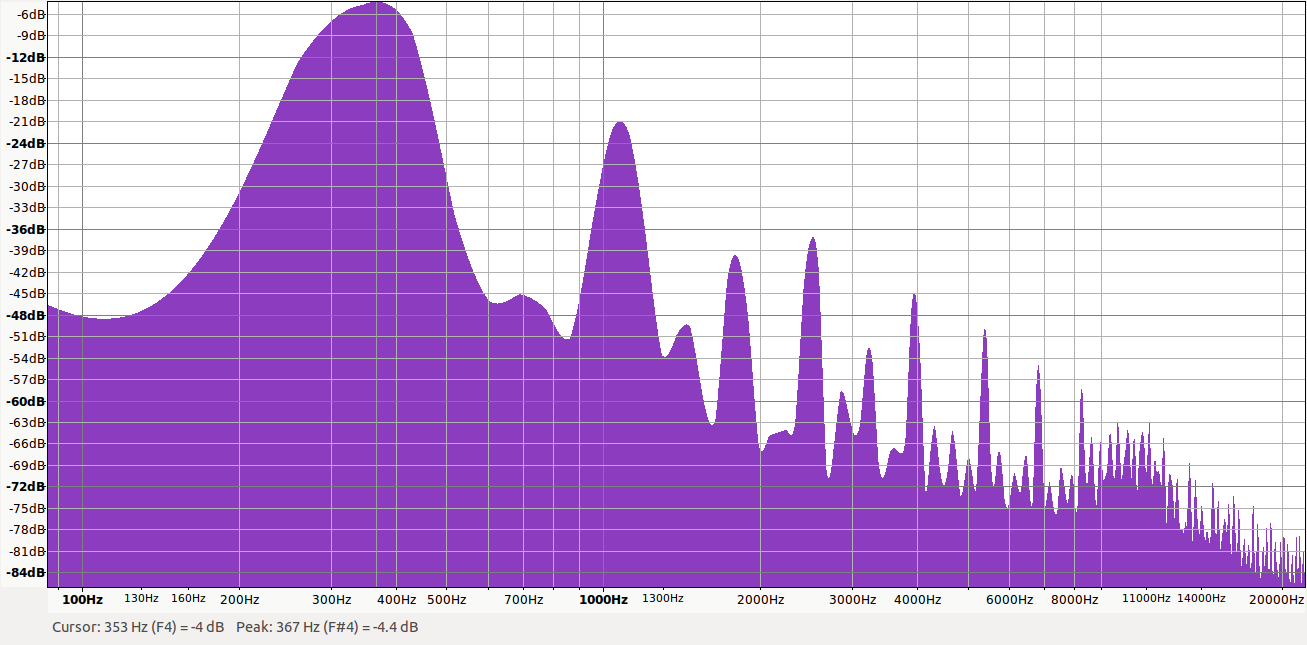}
 \caption{Recorded sound intensity versus frequency for input resonant signal: 367 Hz}\label{f}
\end{figure}

Though unnecessary for the present discussion, it is instructive to look at the Fourier transform that \textit{Audacity} can perform on the sound signal recorded by the microphone when the speaker is fed with a sinusoidal resonant frequency of the tube. Fig.(\ref{f}) shows such a plot where the speaker was fed with a 367Hz signal which was a resonant frequency for a particular length of the tube. 

It can be seen that though the speaker was fed with a single frequency, the constructive interference of certain frequencies contained in the noise are seen as peaks. These are different modes of resonance of the tube at that particular length.

As undergraduate students have more intuition of sound waves than of electronics or computer codes, we expect them to relate more easily to the meaning of Fourier transforms.

\section{Acknowledgement}
We would like to thank Dr. K. Senapati for getting us into this problem. We thank N. Anand for suggesting the Kundt's tube for performing a mechanical FT. We thank Dr. R. Das for his encouragement and School of Physical Sciences - National Institute of Science Education and Research, Bhubaneswar - India  for funding this experiment.

\end{document}